\documentclass[preprint2]{aastex}
\shorttitle{Extrasolar Binary Planets I}
\shortauthors{Ochiai, Nagasawa, \& Ida}

\newcommand{\bun}[2]{\left(\frac{#1}{#2}\right)}
\newcommand{\Vec}[1]{\mathbf{#1}}

\begin{document}

\title{Extrasolar Binary Planets I:\\
 Formation by tidal capture during planet-planet scattering}

\author{H. Ochiai$^1$, M. Nagasawa$^2$, and S. Ida$^3$}
\affil{1) Earth and Planetary Sciences, Tokyo Institute of Technology, 2-12-1 Ookayama, \\
Meguro-ku, Tokyo 152-8551, Japan}
\affil{2) Interactive Research Center of Science,
Tokyo Institute of Technology, 2-12-1, Ookayama, \\
Meguro-ku, Tokyo 152-8551, Japan
}
\affil{3) Earth-Life Science Institute, Tokyo Institute of Technology, 2-12-1 Ookayama, \\
Meguro-ku, Tokyo 152-8550, Japan}


\email{nagasawa.m.ad@m.titech.ac.jp}

\begin{abstract}
We have investigated i) the formation of gravitationally bounded pairs of 
gas-giant planets (which we call "binary planets") from capturing each other 
through planet-planet {\it dynamical} tide during their close encounters
and ii) the following long-term orbital evolution due to
planet-planet and planet-star {\it quasi-static} tides.
For the initial evolution in phase i), we carried out N-body simulations of the systems consisting 
of three jupiter-mass planets
 taking into account the dynamical tide.
The formation rate of the binary planets is as much as 10 \% of the systems 
 that undergo orbital crossing
and this fraction is almost independent of the initial 
stellarcentric semi-major axes of the planets,
while ejection and merging rates sensitively depend on the semi-major axes. 
 As a result of circularization by the planet-planet dynamical tide,
typical binary separations are a few times the sum of
the physical radii of the planets.
 After the orbital circularization, the evolution of the binary system is governed by long-term quasi-static tide.
We analytically calculated the quasi-static tidal evolution in later phase ii).
The binary planets first
enter the spin-orbit synchronous state by the planet-planet tide.
The planet-star tide removes angular momentum of 
the binary motion, eventually resulting in a collision between the planets.
However, we found that the binary planets survive the tidal decay 
 for main-sequence life time of solar-type stars ($\sim 10$Gyrs),
if the binary planets are beyond $\sim0.3$ AU from the central stars.
These results suggest that the binary planets can be detected 
by transit observations at $\ga 0.3$AU.
\end{abstract}


\section{INTRODUCTION}
About a half of the discovered extrasolar gas-giant planets have eccentric orbits with eccentricity $e \ga 0.2$.
Planet-planet scattering is one of the relevant mechanisms to excite their eccentricities
after the formation of the gas giants.
Early studies on this subject \citep[e.g.,][]{RF96,Weiden96} assumed
two gas giants with initially close enough orbital separations for them to start
orbital crossing quickly.
However, it may not be easy to realize such orbitally unstable orbital configurations
as a result of their formation processes. 
On the other hand, systems of three gas giants with modest orbital separations
start orbital crossing as a result of long-term secular perturbations
well after the formation of the systems,
such systems have been regarded as more plausible
initial conditions and they have been intensively studied by later papers
\citep[e.g.,][]{LI97,MW02,c08,Juric08}.
During the orbital crossing, the planets repeatedly undergo close encounters
and their eccentricities are highly pumped up.
Typical fates of the three planet systems are ejections of a planet, 
planet-planet collisions, and planet-star collisions.
Through these events, usually the remaining two planets acquire widely separated eccentric orbits
and further orbital crossing does not occur.

In the simulations that include planet-star tidal interactions,
the gravitational scatterings can lead to formation of hot jupiters.
If the eccentricity of a planet is excited up to $\sim 1$,
the pericenter becomes very close to the host star.
Then, the tidal dissipation 
of the planet induced by the star (dynamical tide) damps its semi-major axis and eccentricity 
and the planet becomes a hot jupiter \citep{RF96}.
\citet{nagasawa} and \citet{BN12} investigated the orbital evolution
of the three gas-giant planets systems that initially have circular orbits 
beyond the snow line,
by N-body simulations including the planet-star tidal interaction.
 Although most of the hot jupiters fall inside of the stellar Roche radius 
due to subsequent tidal decay during timescales 
of the order of 1 Gyr \citep{BN12}, the previous papers found that hot jupiters are formed 
in as much as 10-30\% of the systems that undergo orbital crossing.

The tidal dissipation also occurs between closely encountering two planets.
So, there is another possibility of the fate of planet-planet scattering, 
that is, formation of binary planets.
This possibility has not been studied in the planet-planet scattering scenarios.
\citet{posi} studied the formation of binary planets in extrasolar planetary systems
for the first time 
and demonstrated that binary planets can be actually
formed from two gas giant planets in 1--20\% of the runs at 0.2--5AU,
by orbital calculations including planet-planet tidal interactions.

However, \citet{posi} started calculations from 
two gas giant planets in almost circular orbits 
with the orbital separation between the two planets 
of $\sim 2.4r_{\rm Hill}$, where $r_{\rm Hill}$ is the Hill radius.
As mentioned in the above, these unstable orbital configurations would not
be established in real systems, and
this initial condition may make trapping of binary planets easier.
With this conditions,
the planets immediately undergo close encounters 
before their eccentricities are pumped up.
As a result, the close encounters have 
relatively low relative velocity in this case,
and the trapping probability is higher for lower relative velocity.

Dynamical behaviors in two planet systems are qualitatively different from 
that with three planets or more.
While close encounters cannot occur in the two planet systems
unless their orbital separation is smaller than $2\sqrt3r_{\rm Hill}$ \citep{Gladman},
the three planet systems do not have such a solid stability boundary.
With modest initial orbital separations, the three planet systems can
start orbital crossing after their eccentricities are built up 
on relatively long timescales \citep[e.g.,][]{Chambers}.
Note that such dynamical behavior is similar even if the number of planets
is increased from three.
 Although it is not clear how much fraction of planetary systems actually 
become unstable after the gas disk dispersal (e.g., Lega et al. 2013), 
calculations starting
from three planets in modestly separated orbits
would be much more appropriate than that from two planets in packed
circular orbits, in order to evaluate the formation rate of binary gas giant planets
in extrasolar planetary systems.
In a separate paper \citep[][which we refer to Paper II]{Ochiai}, we will discuss the
detectability of binary planets by transit observations.
For such discussions, statistical results
based on simulations of planet-planet scattering 
of three planets are more helpful than those with two planets.

In this paper, we carry out N-body simulations of systems of three giant planets,
taking account of planet-planet interactions by dynamical tide as well as planet-star ones to simulate tidal capture of the planets with each other and circularization of the captured orbits. 
In \S \ref{keisan}, we describe basic equations of the N-body simulation 
and the model for the dynamical tide that is incorporated in the N-body simulation.
In \S \ref{kekka}, we present the numerical results.
We show that the formation rate of the binary planets is as high as 
$\sim10$ \% almost independent of stellarcentric semimajor axis in the three planet system.
 After orbital circularization due to dynamical tide,
the evolution of the binary system is governed by long-term quasi-static tide.
We analytically calculate the long-term tidal orbital evolution
of the formed binary planets during
main-sequence phase lifetimes of central stars.
Section \ref{kousatsu} is the conclusions.

\section{METHODS}
\label{keisan}
\subsection{Basic equations of N-body simulation}

We consider the planetary systems, which are composed of a central star and 
three gas giant planets.
We take the origin of the coordinate at the central star with mass $M_\ast$.
The equation of motion of planet $i$ is

\begin{eqnarray}
\frac{d^2\Vec r_i}{dt^2}=&&\!\!\!\!\!\!\!\!\!-G\frac{M_\ast+M_i}{r_{i}^3}\Vec r_{i}-GM_j\left(\frac{\Vec r_j}{r_{j}^3}+\frac{\Vec r_{ij}}{r_{ij}^3}\right)\nonumber\\
&&\!\!\!\!\!\!\!\!\!-GM_k\left(\frac{\Vec r_k}{r_{k}^3}+\frac{\Vec r_{ik}}{r_{ik}^3}\right) ,
\end{eqnarray}
where $M_i$ and $\Vec r_i$ are masses and position vectors of planet $i$ (= 1, 2, and 3), respectively, and $\Vec r_{ij}\equiv \Vec r_i-\Vec r_j$.

When a planet passes the pericenter to the star or another planet,
we impulsively dissipate the orbital energy of the passing bodies
according to tidal interactions, following \citet{NI11}.
We impose the tidal interactions with the host star 
only for encounters with the stellarcentric pericenter distance $q < 0.04$ AU
to save computational time.

When a planet repeatedly undergoes close encounters with the central star
and suffers the tidal dissipation,
its semimajor axis and orbital eccentricity shrink
keeping $q$ almost constant \citep[e.g.,][]{nagasawa,NI11}.
As a result, the planet becomes a hot jupiter.

If a planet encounters another planet closely enough,
the planets can be trapped through planet-planet tidal dissipation
to form a gravitationally bound pair.
At the initial phase after the trapping, 
eccentricity of the binary orbit is generally close to unity.
Through repeated encounters, the binary orbit
is circularized in a similar way to the formation of hot jupiters.  
 Since the relative motion is described by hyperbolic orbits at the trapping,
and by highly eccentric orbits in most of time during the circularization,
the tidal interaction is dominated by dynamical tide, which is described in the following.

\subsection{Tidal dissipation between planets}
\label{chousekironbun}

 For tidal trapping and circularization of binary orbits,
we use the formula for energy loss due to dissipation due to dynamical tide between two objects
derived by \citet{zm}, following \citet{posi}.
When planet $i$ and $j$ undergo tidal interactions, 
the tidal energy loss caused by a single close encounter between the planets is
\begin{eqnarray}
E_{{\rm tide}}
 =&&\!\!\!\!\!\!\!\!\!\!\!\! \frac{GM_j^2}{R_i}\left[\bun {R_i}{q_{ij}}^6 \!\! T_2(\eta_i) \! +\! \bun {R_i}{q_{ij}}^8 \!\! T_3(\eta_i)\right] \nonumber \\
&& \hspace{-1.5cm} + \frac{GM_i^2}{R_j}\left[\bun {R_j}{q_{ij}}^6 \!\!T_2(\eta_j)\! +\! \bun {R_j}{q_{ij}}^8 \!\! T_3(\eta_j)\right],
\label{ZMdE}
\end{eqnarray}
where $q_{ij}$ is the pericenter distance between planet $i$ and $j$,
$R_i$ is planetary physical radius, $\eta_i \equiv \{M_i/(M_i+M_j)\}^{1/2}(q_{i,j}/R_i)^{3/2}$, and
$T_{2,3}(\eta_i)$ are given by \citet{zm} as a fifth-degree polynomial function.

For encounters between the planets with the pericenter distance 
less than $10(R_i+R_j)$,
we subtract tidal dissipation energy from their orbital energy
at the pericenter passage using impulse approximation.
Because tidal interactions rapidly weaken as the distance between the bodies increases,
even if we switch on the tidal interaction at larger distance,
the results hardly change (see \S \ref{keiseijouken}).

Strictly speaking, it may not be correct to use the impulse approximation and equation (\ref{ZMdE}) after the eccentricity of the binary orbit is significantly damped,
because the oscillation of the planetary bodies raised by the tides
is not dissipated before the next encounter.
In such situations, how the semi-major axis and eccentricity 
of the binary orbit are changed by dynamical tides depends 
on the phase of oscillation, and $a$ and $e$ change chaotically \citep[e.g.,][]{PT77,Mardling}.
However, we are most concerned with early phase of the tidal trapping
to evaluate the formation probability of binary planets.
So, we neglect the chaotic evolution due to incomplete oscillation damping.

 Note that the formulas for dynamical tide could include large uncertainty, because the effect of dynamical tide depends on what wave modes are excited and how they dissipate. 
However, the tidal dissipation energy is inversely proportional to several powers of mutual distance between the planets. 
In the case of equation (\ref{ZMdE}), $E_{\rm tide} \propto q_{ij}^{-6}$. 
Even if $E_{\rm tide}$ cahnges by a factor 10, the capture
separation changes only by 50\%.

\begin{table*}[tbh]\begin{center}
\begin{tabular}{cccccccc}\tableline\tableline
case     &$a_{1}$(AU) & $R_i$ ($R_{\rm J}$) & binary planets & collision & HJs & ejection & 3 remain \\ \tableline
set 1    &1 & 2 & 8 & 68 & 8 & 16 & 0  \\
set 3    &3 &2 & 10 & 30 & 17 & 41 & 2 \\
set 5    &5 & 2 & 9 & 32 & 20 & 38 & 1 \\
set 10   &10 &2 & 13 & 15 & 16 & 54 & 2 \\ \tableline
set 0.5a &0.5 & 1& 9 & 64 & 12 & 15 & 0 \\
set 3a   &3 & 2,1,1 & 17 & 23 & 23 & 37 & 0 \\
set 3b   &3 & 2 & 6 & 41 & 15 & 33 & 0 \\
set 5a   &5 & 2  & 8 & 34 & 16 & 42 & 0 \\ \tableline
\end{tabular}
\caption{
The parameters and results of 8 sets of N-body simulations
of three gas giant planets.
Each set includes 100 runs with different initial orbital phases.
The parameter $a_1$ is the initial stellarcentric semimajor axis
of the innermost planet and $R_i$ is the physical radius of the planets
in unit of Jovian radius $R_{\rm J}$.
The rows "binary planets", "collision", "HJs", and "ejection"
refer to the number of runs that end up with formation binary planets,
planet-planet and planet-star collisions, formation of hot jupiters
through tidal circularization, and ejection from the systems, respectively.
The last row represents the number of runs in which orbital interaction
of three planets continues until the end of calculations (10 Myrs). 
\label{100toorinokekka}}\end{center}
\end{table*}

After the circularization, quasi-static tides become predominant instead of dynamical tides. Although quasi-static tides are weaker by orders of magnitude than dynamical tides for high orbital eccentricity, quasi-static tides work also for circular binary orbits and last until spin-orbit synchronous state is established. Thus the cumulative effect in orbital evolution due to quasi-static tide is important when we consider observational detectability of binary planets.
In \S \ref{chousekishinkanosetsu}, we calculate such longer-term tidal evolution due to quasi-static tides between planets, taking into account planetary spins and also tides between the planets and the host star.
The formulation is based on tidal dissipation functions $Q$, instead of eq.~(\ref{ZMdE}), as explained in Appendix A.
We discuss orbital stability against tidal evolution due to quasi-static tides during main sequence lifetime of solar-type host stars ($\sim$ 10 Gyrs).  We will show that a binary system is stable if stellarcentric distance is $\ga$ 0.3-0.4 AU.  Also note that as long as the tidal capture occurs beyond $\sim$ 0.3AU from the central star, envelope removal due to the tidal dissipation do not take place (see \S \ref{keiseijouken}).

\subsection{Initial conditions}
\label{shokijouken}
To evaluate the formation rate of the binary planets,
we set three giant planets.
In early studies, planet-planet scattering simulations 
started from two planet systems,
but recent simulations are all done using three planets or more \citep[e.g.,][]{BN12}.

However, since the pioneering work, \citet{posi}, 
carried out two planet simulations 
 as one set of runs,
we also performed the same two planet simulations to confirm
that our simulation code reproduces the \citet{posi}'s results.
For each semimajor axis, we carried out 20-30 runs.
We found that in the {\it two planet} cases,
the formation rate of binary planets are 1 out of 20 at 0.2 AU, 5 out of 20 at 1.0 AU, and 11 out of 30 at 5.0 AU, respectively.
The results are consistent with \citet{posi}'s,  
although the numbers of our runs are smaller than theirs.

\begin{figure}[thb]
\includegraphics[width=7cm]{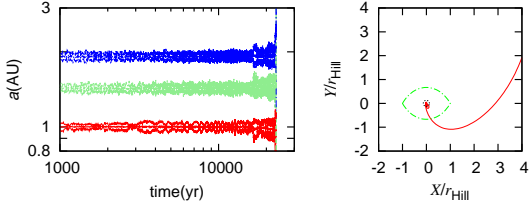}
\caption{
An example of orbital evolution to form binary planets.
We set $a_1 = 1$AU.
The left panel represents the time evolutions of the semi-major axes.
The solid red line, dashed light green line, and dash-dotted blue line are planet 1, 2, and 3, respectively.
Two thin lines mean the pericenter distance $a_i(1-e_i)$ and apocenter distance $a_i(1+e_i)$.
The right panel represents the relative orbit in the Hill coordinate (the solid red line) between planet 2 and 3.
The dash-dotted green line and dotted black line indicate the Hill sphere and 
the region inside which tidal interactions are taken into account, respectively.
$X$ and $Y$ are radial and tangential coordinates normalized by the Hill radius.
\label{f1}
}
\end{figure}

Our main results in this paper are obtained by {\it three planet} simulations.
We use initial conditions as follows:
the semi-major axes of three planets are $1.0a_1$, $1.45a_1$, and $1.9a_1$ and we test four different initial semi-major axis of the innermost planet, $a_1=1, 3, 5,$ and 10 AU.
Initial orbital eccentricities are zero for all the planets,
but we set small orbital inclinations as $I_1=0.5^{\circ},I_2=1.0^{\circ}$, 
and $I_3=1.5^{\circ}$ to ensure three-dimensional motions, 
following \citet{MW02}.
We set the other angle variables (the longitudes of ascending node, those of pericenter, and the mean longitude) randomly.
For different runs with the same $a_1$ and planetary radii ($R_j$), we use different seeds for the random number generation.  
The mass and radius of the central star are $1M_{\odot}$ and $1R_{\odot}$, and that of three planets are $1M_{{\rm J}}$ and $2R_{{\rm J}}$, respectively.
Since the orbital crossing and formation of binary planets may occur at the timings 
before the envelope of gas giants fully contract,
we use relatively large $R_i$.
We also performed additional set of calculations with $a_1=0.5$ AU and $R_j = 1R_{{\rm J}}$.
 Unless we particularly note that $R_j = 1R_{{\rm J}}$,
we use $R_j = 2R_{{\rm J}}$ as a nominal parameter (also see Table~\ref{100toorinokekka}).

For each $a_1$, we carried out 100 runs to follow the orbital evolution over 10 Myrs
with 4-th order Hermite scheme.
We stopped calculations when a pair of planets collide ($r_{ij}\le R_i+R_j$)
or when they form a binary system and the binary eccentricity $e_{\rm bi}<0.01$).
We regard that a planet is ejected from the system when 
the instantaneous distance of the planet from the star 
becomes more than 10000 AU and planet's eccentricity exceeds unity.
We also check the collision between the central star and a planet,
but when a planet approaches the central star, it tends to become a hot jupiter due to the tidal interaction with the central star rather than a collision.
We neglect spins of the planets and the central star 
in the N-body simulations for the simplicity
(The spin evolution is calculated in the long-term evolution
due to quasi-static tide shown in section \ref{chousekishinkanosetsu}).

\begin{figure}[tbh]
\includegraphics[width=8cm]{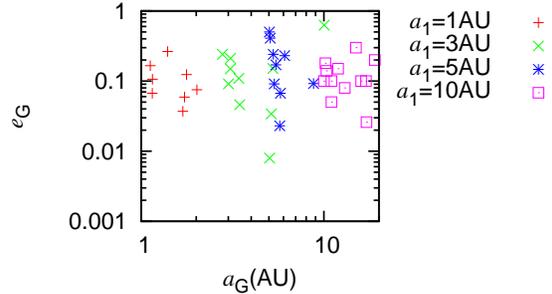}
\caption{
The stellarcentric semi-major axes and eccentricities of binary's centers of mass of four initial $a_1=1$ AU (red plus), 3 AU (green cross), 5 AU (blue asterisk), and 10 AU (magenta square).
\label{aeg}
}
\end{figure}

\section{RESULTS}
\label{kekka}
\subsection{Distributions of orbital parameters of formed binaries}

Figure~\ref{f1} shows an example of orbital evolution to form a binary of planets.
The left panel shows the time evolution of the semi-major axes of the
three planets.
Two thin lines mean the pericenter distance $a_i(1-e_i)$ and apocenter distance $a_i(1+e_i)$.
Orbital crossing begins at $t\sim23000$ yr.
Immediately after that,
planet 2 (dashed light green line) and 3 (dash-dotted blue line) form a binary.

\begin{figure}[htbp]
\epsscale{0.7}
\plotone{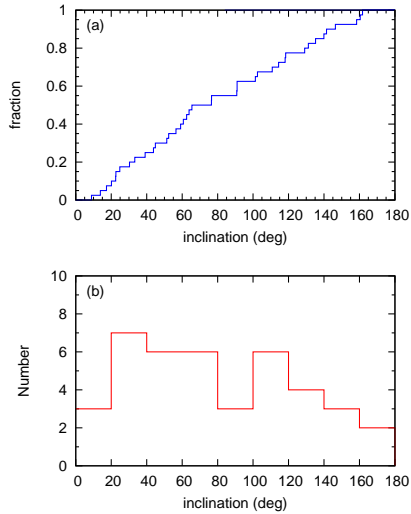}
\caption{
The stellarcentric orbital inclinations of binary planets $I_{\rm bi}$ obtained in 400 calculations.
This is sum of four initial $a_1$ in $2R_{\rm J}$ cases. (a) Cumulative inclination distribution. (b) Histogram of inclination with stride of 20 degrees.
\label{Ibi}
}
\end{figure}

The right panel represents the path of approaching planets
in the local Hill coordinate.
The inner planet (planet 2) is set at the origin.
In this panel, $X$ axis lies along the line from the central star to the inner planet
and $Y$ axis is perpendicular to $X$.
The axes are normalized by Hill radius defined by
\begin{eqnarray}
r_{\rm Hill} = \bun{M_i+M_j}{3M_\ast}^{1/3}\frac{M_ia_i+M_ja_j}{M_i+M_j}.
\end{eqnarray}
The outer planet (denoted by the solid red line) enters the Hill sphere (denoted by the dash-dotted green line) at $t\sim23000$ yrs.
After that two planets undergo repeated close encounters to suffer tidal interactions (dotted black line
indicates the region within which tidal interaction is taken into account) 
for $\sim300$ yrs, they form a gravitationally bound pair (binary).
We found that
when the planet enters the Hill sphere from upper-left (lower-left) direction in this plot,
the planets tend to form a prograde (retrograde) binary.

We present the distribution of the stellarcentric semi-major axes and the eccentricities of binary's barycenters ($a_{\rm G}$ and $e_{\rm G}$) in Fig. \ref{aeg}.
Four different symbols represent the different initial semi-major axes $a_1=1$ AU (plus), 3 AU (cross), 5 AU (asterisk), and 10 AU (square).
This figure shows that the binary planets are formed near their initial orbits.
This is because the binary planets are formed in the early stage of orbital instability 
before they are significantly diffused by many scatterings.
The (stellarcentric) eccentricities of the barycenters of the binary pairs are distributed in 
the range of 0.01--0.6 with the mean value $\sim 0.15$.
The mean value corresponds to the eccentricity
required for close encounters from initial orbital separation of planets $\sim 4r_{\rm Hill}$. 
Since the resultant eccentricities $e_{\rm G}$ depend on the phase angle of the encounters,
they are distributed in a broad range.

\begin{figure}[tbh]
\includegraphics[width=7cm]{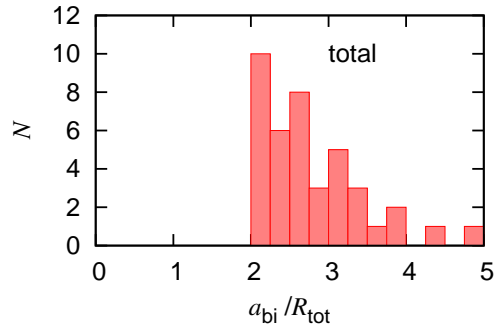}
\caption{
The semi-major axis of binary orbits $a_{\rm bi}$ (distance between binaries) obtained in 400 calculations.
This is sum of four initial $a_1$ in $2R_{\rm J}$ cases.
\label{abinary}
}
\end{figure}

Figure~\ref{Ibi} shows the stellarcentric orbital inclination of formed binary planets, $I_{\rm bi}$$=$$\arccos(h_{ij,Z}/h_{ij})$,
where $h_{ij,Z}$ is the component of $\Vec h_{ij}=\Vec r_{ij} \times \Vec v_{ij}$ that is perpendicular to the orbital plane, and
$\Vec v_{ij}$ is the relative velocity between planet $i$ and $j$.
Because we do not see any significant semimajor axis dependence of $I_{\rm bi}$,
we superpose the results from different initial $a_1$.
If tidal trapping occurs isotropically, $I_{\rm bi}$ becomes a sine distribution.
 KS test suggests that this distribution is not the isotropic distribution.
The distribution is slightly skewed to small $I_{\rm bi}$,
which might reflect the fact that the trapping occurs
in early phase before significant orbital excitations.
But, retrograde binary planets are formed in non-negligible fraction of runs (18/40).
Note that since we do not take into account the spins of the planets, the retrograde orbits do not necessarily mean the tidally unstable configuration.

Figure~\ref{abinary} shows the final semi-major axis of the binary orbits, $a_{\rm bi}$.
Because we plot the values after the binary eccentricities have been significantly damped
($e_{\rm bi} < 0.01$),
$a_{\rm bi}$ is equivalent to the binary separation.
Here we also superpose the results from different initial stellarcentric $a_1$.
The distribution of $a_{\rm bi}$ is peaked at $2R_{\rm tot}-4R_{\rm tot}$
where $R_{\rm tot}=R_i+R_j$.
This is about twice as large as the pericenter distances 
just after binary capture (see \S \ref{keiseijouken}).
The factor of 2 is attributed to angular momentum conservation
during the tidal circularization as below.

\begin{figure}[tbh]
\includegraphics[width=7cm]{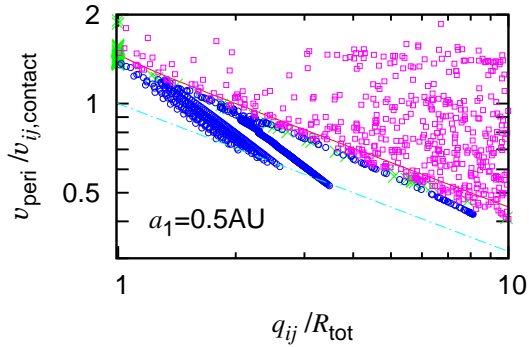}
\caption{
The relative distances and velocities at the pericenters of two encountering planets of four initial semi-major axes.
The velocities are normalized by the relative velocity of the contact binary.
The solid red lines and the dash-dotted light blue lines represent the escape velocity and the Keplerian velocity of the binary planet, respectively.
The blue open circles represent pericenter passages of planets that become binaries in the end.
Open magenta squares and green crosses show pericenter passages of escaping pairs and colliding pairs, respectively.
\label{dv}
}
\end{figure}
Because tidal force is a strong function of a separation distance between
two planets, the two planets just after the trapping usually have 
a highly eccentric binary orbit with $q_{\rm bi,0} \sim R_{\rm tot}-2R_{\rm tot}$.
The initial angular momentum of the binary orbit is
$h_{\rm bi,0} \simeq [a_{\rm bi,0}(1-e_{\rm bi,0}^2)]^{1/2}\simeq (2q_{\rm bi,0})^{1/2}$,
while the final angular momentum is $h_{\rm bi} \simeq a_{\rm bi}^{1/2}$.
The angular momentum conservation indicates that $a_{\rm bi} \sim 2q_{\rm bi,0}$.
This explains the peak at $2\la a_{\rm bi}/R_{\rm tot}\la 4$ and deficit
of binary planets at $1 \la a_{\rm bi}/R_{\rm tot} \la 2$ in Fig.~\ref{abinary}.
We found that the peaked value of $a_{\rm bi}$ is much smaller than the Hill radius $\sim 40R_{\rm tot}$ at 1AU, and the distribution of 
$a_{\rm bi}$ does not depend on $a_1$.
This indicates that the binary is hardly affected
by interactions with the star or a third planet outside Hill sphere,
unless $a_{\rm G}$ is very small.

%
%
%
%

\subsection{Tidal capture and orbital circularization due to dynamical tide}
\label{keiseijouken}

In this subsection, we show the process of tidal trapping in more details. 
We assumed that the relative velocity of the planets is impulsively decreased at their closest approach.
We expect that
the binary planets are formed when the relative velocity ($v_{ij}$) of two planets immediately 
after the impulsive tidal dissipation is smaller than their escape velocity,
\begin{eqnarray}
v_{ij} < v_{\rm{esc}}=\sqrt{\frac{2G(M_i+M_j)}{r_{ij}}}.
\label{vesc}
\end{eqnarray}
In Fig. \ref{dv}, we present the relative distances ($q_{ij}$) and 
the relative velocities ($v_{\rm peri}$) 
after the tidal dissipation
of two planets at each pericenter passage,
in the case of $a_1=0.5$ AU with $R_{i,j}=1R_{\rm J}$.
The velocities are normalized by the relative velocity of the contact binary, $v_{ij,\rm{contact}}=[G(M_i+M_j)/R_{\rm tot}]^{1/2}$,
and the pericenter distances are normalized by $R_{\rm tot}$.
When $q_{ij}/R_{\rm tot}$ becomes less than unity, two planets collide.
The solid red line represents $v_{\rm peri}= v_{\rm{esc}}$.
The passages below this line correspond to those of bound orbits.
The dash-dotted (light blue) lines represent the binary's circular orbital velocity $v_{\rm K,bi}=[G(M_i+M_j)/r_{ij}]^{1/2}$.
When $v_{\rm peri}$ reaches this line, their tidal circularization 
have been completed.
Open blue circles, magenta squares and green crosses represent 
a series of the pericenter passages of binary, escaping, and colliding pairs, respectively.
At points along the line of $v_{\rm peri}= v_{\rm{esc}}$, a binary 
(a gravitationally bound pair) is first formed.
As the binary bodies repeat close approaches,
$v_{\rm peri}$ is decreased by tidal dissipation and $q_{ij}$
is increased by the angular momentum conservation, which
are represented by a chain of blue open circles 
from upper-left to lower-right direction in the figure. 
We found that the first tidal captures of binary planets occur
when $q_{ij} \sim R_{\rm tot}-2R_{\rm tot}$,
that is, $q_{ij}$ is small enough for the tidal force to be strong enough
but larger than $R_{\rm tot}$ to avoid a collision.
In all of our simulations, no tidal capture was found 
from non-bound orbits with $q_{ij}> 3R_{\rm tot}$ 
(The blue open circles outside of $4R_{\rm tot}$ in the figure
are wondering passages during the circularization that
start from $q_{ij} \sim R_{\rm tot}-2R_{\rm tot}$).

As already mentioned, 
we included the tidal dissipation
when the closest approach occurs within $10R_{\rm tot}$.
As a stellarcentric distance of binary planets increases, the Hill radius becomes larger, but we did not change the threshold distance of $10R_{\rm tot}$.
We carried out extra calculations at $a_1=5$ AU with
the threshold distance of $50R_{\rm tot}$, to check its effect (set 5a in Table \ref{100toorinokekka}).
The formation rate of binary planets is 8 \% in this calculations, and
binary formation occurred at $q_{ij}>10R_{\rm tot}$ in only 1 of 100 runs.
So, the results hardly change even if the tidal force is incorporated
from more distant encounters than $q_{ij}=10R_{\rm tot}$.

 Note that tidal destruction of planets hardly occurs in this capture process.
The relative velocity at grazing approach ($q_{ij} \sim R_{\rm tot}-2 R_{\rm tot}$)
that causes tidal capture, is $v \simeq [ v_{\rm gr}^2 + (e v_{\rm Kep})^2]^{1/2}$, where 
$e v_{\rm Kep}$ is relative velocity when the planets are sufficiently separated
and $v_{\rm gr}$ is a contribution by the gravitational acceleration between interacting planets, which is given by $\sim (0.5 - 1)v_{\rm esc}$ for $q_{ij} \sim 2R_{\rm tot}-1R_{\rm tot}$.
For nominal parameters, $1M_{\rm J}$ and $2R_{\rm J}$, the surface escape velocity
is $v_{\rm esc}\simeq 44$km/s.
The Keplerian velocity is $v_{\rm Kep}\simeq 30(a/1{\rm AU})^{-1/2}$km/s.
Since typical stellarcentric eccentricity is $e \lesssim 0.3$ at the capture (Fig. \ref{abinary}), $v_{\rm gr}$ is dominated and 
$v \sim (0.5-1)v_{\rm esc}$ for $a \ga 0.3$AU, where the orbits of binary planets are stable on timescales of $\sim 10$ Gyrs (see section~\ref{chousekishinkanosetsu}).
For collision cases, SPH simulation \citep[e.g.,][]{Iko06} showed that significant envelope loss occurs only for $v \ga 2 v_{\rm esc}$. 
Therefore, the tidal destruction is unlikely.

The tidal dissipation could inflate the planetary envelope, which
accelerates tidal circularization.
However, while this may change total circularization timescale, it would not significantly change binary orbital separations after the circularization, because 
the separations are regulated by $q_{ij}$ at the trapping when the inflation has not been caused.

\subsection{Formation rate of binary planets}
\label{formation rate}

\begin{figure}[tbh]
\includegraphics[width=7cm]{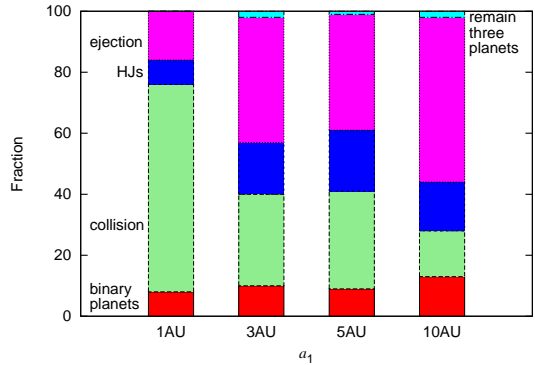}
\caption{
The results of 400 runs for four kinds of initial semi-major axes $a_1=1$, 3, 5, and 10 AU. The colors represent binary planets (red), collision (light green), hot jupiters (blue), ejection (magenta), and remain three planets (light blue).
\label{100kekka}
}
\end{figure}

We summarize the results of four sets of 100 runs with
initial stellarcentric semi-major axes $a_1=1$, 3, 5, and 10 AU
(set 1, 3, 5 and 10)
 for $R_i=2R_{\rm J}$,
in Fig. \ref{100kekka} and Table \ref{100toorinokekka}.
The ejection rate increases and the collision one decreases as
stellarcentric semimajor axis increases, because $R_i/r_{\rm Hill}$ decreases with
the semimajor axis.
The important result is that the formation rate
of the binary planets is $\sim10$ \%, almost independent of the semimajor axis
(in other words, almost independent of 
the value of $R_i/r_{\rm Hill}$) as long as $a_1=1$--10 AU.

Compared with the results of two-planet systems, 
the binary formation rate is lower, because 
the initial conditions of two planet simulations 
cause tidal capture before stellarcentric eccentricity is excited. 
In the three planet cases, the orbital behaviors leading to tidal capture
is much more complicated than in the the two planet systems
and eccentricity is excited enough before the capture that the frequency of close encounters with relatively low relative velocity is diminished.
Nevertheless, the formation probability
of the binary planets in three planet systems is still as large as $\sim10$ \%.

In order to check how the formation rate
of the binary planets depends on other parameters,
we carried out four additional calculations 
(set 0.5a, 3a, 3b, and 5a in Table~\ref{100toorinokekka}).
In these additional calculations,
we changed one of the parameters (planetary radius, initial planet-planet distances, or the tidal limit) and keep the other parameters including
planetary masses the same.
In set 0.5a, we used half-sized planets ($R_i = 1R_{\rm J}$),
keeping the planetary masses the same.
 Since $a_1=0.5$AU, this set has the same value of $R_i/r_{\rm Hill}$ as set 1.
As a result, the ejection/collision ratio is similar, as shown in Table~\ref{100toorinokekka}.
Although the effect of tidal dissipation is weaker than in set 1,
we found that a similar $(9/100)$ binary formation rate. 
The binary planets are formed through grazing encounters. 
When we use the smaller planetary radius, while the tidal dissipation becomes weaker,  
some fraction of the close encounters that lead to collisions for $R_i = 2R_{\rm J}$
result in binary formation.
These two effects tend to cancel, so that
we did not find a large difference between set 1 and set 05a 
in our small number of statistics.

In set 3a $(a_1 = 3$AU), while we keep $R_1 = 2R_{\rm J}$, $R_2$ and $R_3$ were reduced to $1R_{\rm J}$.
The distribution of semi-major axes of half-sized binary planets is peaked at
$a_{\rm bi}\sim 2.5R_{\rm tot}$, which is similar to Fig.~\ref{abinary}.
However, the formation rate of the binary planets is increased to be $\sim 17\%$.
The increase comes from formation of $2R_{\rm J}-1R_{\rm J}$ binaries (11\% of the 17\%).
This may be because the $2R_{\rm J}$ can suffer stronger tidal forces without collisions.

In set 3b $(a_1 = 3$AU), the planet-planet initial orbital separation set to be
$\sim6r_{\rm{Hill}}$ at $a_1=3$ AU, which is 1.5 times
larger than the standard cases.
Table \ref{100toorinokekka} shows that the formation probability
of binary planets is slightly lower ($\sim 6\%$).
This is because the planets have to build up larger eccentricities 
for close encounters to occur and accordingly the relative velocities are higher.


In set 5a $(a_1 = 5$AU), we increased the threshold distance inside which the tidal force is incorporated,
by five times at $a_1=5$ AU.
The formation probability of the binary planets does not change,
as we mentioned in \S \ref{keiseijouken}.

\subsection{The stability of binary planets against long-term evolution due to stellar and planetary quasi-static tides}
\label{chousekishinkanosetsu}

\begin{figure}[tbh]
\includegraphics[width=.4\textwidth]{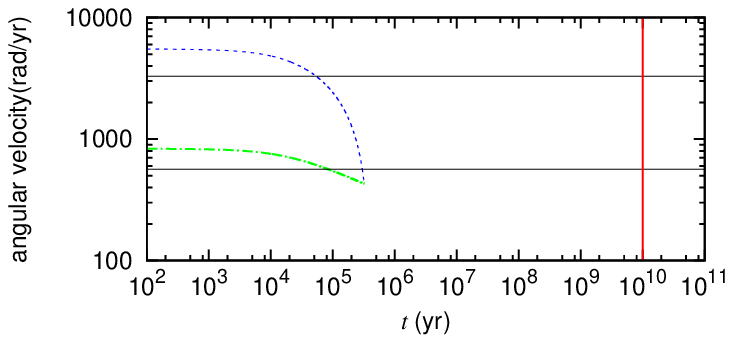}

\includegraphics[width=.4\textwidth]{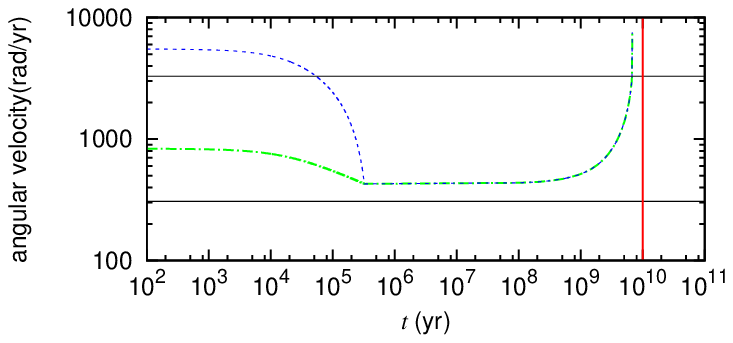}

\includegraphics[width=.4\textwidth]{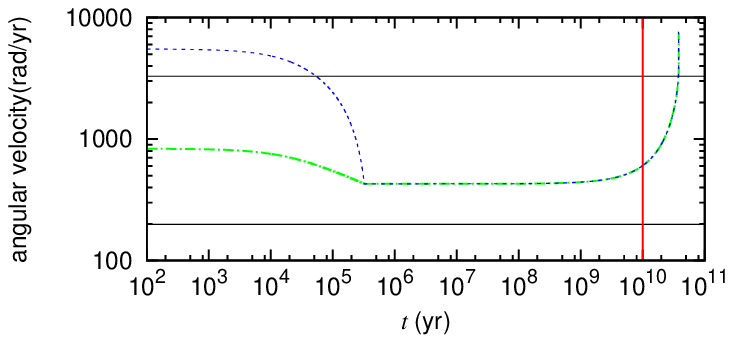}
\caption{
The orbital evolutions of the binary planet at $a_{\rm G}=0.2$ AU (upper panel), 0.3 AU (middle panel), and 0.4 AU (lower panel). 
The dash-dotted green line represents orbital mean motion of the binary planet, the dashed blue line shows the spin rates of planets,
the vertical red lines represent the lifetime of the solar system ($t=10^{10}$ yr),
the upper horizontal black line shows the orbital angular velocity of the contact binary $\sqrt{G(M_{\rm pp}+M_{\rm cp})/(R_{\rm pp}+R_{\rm cp})^3}$, and the lower horizontal black line is the critical orbital angular velocity $n_{\rm crit}$.
\label{jumyou}
}
\end{figure}

In previous sections, we showed that the binary gas giant planets are 
formed through the scattering of three planets orbiting around a central star
and planet-planet dynamical tide with no-negligible probability ($\sim 10\%$).
 After the orbital circularization, the dynamical tide diminishes and
the orbital evolution of the binary system is governed by long-term quasi-static tides.
In this section we calculate long-term tidal evolution of the formed binary systems
through the planet-planet and planet-star {\it quasi-static} tidal interactions,
 instead of dynamical tide that have been considered in previous sections.
The evolution on main-sequence lifetime of solar type stars ($\sim 10$Gyrs) is
very important for detectability of binary planets in extrasolar planetary systems
by, e.g., transit observations (see Paper II).

We calculate this tidal evolution process, basically following \citet{Sasaki}.
\citet{Sasaki} neglected the tidal interactions between 
the central star and the binary companion and their spins.
Because we consider binary planets with relatively small
stellarcentric radius where the detectability by transit observation is not too small,
we include all the tidal interactions and the spin angular momenta 
in the planet-companion-star systems.

Although we consider a pair of comparable
planets,
we call one planet "primary" and the other "companion" for convenience.
We use the subscripts ''$\ast$'' for the central star, ''pp'' for the primary planet, 
and ''cp'' for the companion planet.
In this subsection, we use the following assumptions:
\begin{enumerate}
\item
The total angular momentum, that is, the sum of the angular momenta of the stellar and the planets' spins 
and those of the binary and stellarcentric orbits,
is conserved.

\item
All orbits in the system are circular and coplanar.
\item
All the spin angular momentum vectors are parallel to their orbital angular momentum vectors.
\item
The separation of two objects can be changed only by the tidal interactions between them and those with the host star. 
It is not affected by the other objects.
\item
The masses of the planets are negligible compared with that of the central star, i.e., $M_\ast \gg M_{\rm pp},M_{\rm cp}$.
\end{enumerate}
The parameters we adopt in the calculations are shown in Table \ref{alphak}.
We set $t=0$ at the completion of tidal circularization of the binary.
We assume that the initial planetary and stellar spin periods are 10 hours
and 30 days, respectively.
The binary orbital period is calculated by $a_{\rm{bi}}$;
it is about 3 days for $a_{\rm{bi}}=2.5R_{\rm tot}$.
The stellarcentric orbital period is calculated by
its semimajor axis of the binary barycenter;
it is 1 year for $a_{\rm G}=1$ AU.
Thus, it is reasonable to assume that 
$\Omega_{\rm pp}=\Omega_{\rm cp}>n_{\rm{bi}}>n_{\rm G}$ at $t=0$,
where $\Omega_{\rm pp}$ and $\Omega_{\rm cp}$ are the planetary spin frequency,
$n_{\rm{bi}}$ and $n_{\rm G}$ are mean motions of
binary and stellarcentric orbits, which are given by
\begin{eqnarray}
\label{n_bi}
n_{\rm bi} &=& \sqrt{\frac{G(M_{\rm pp}+M_{\rm cp})}{a_{\rm bi}^3}}, \\
n_{\rm G} &=& \sqrt{\frac{G(M_\ast + M_{\rm pp}+M_{\rm cp})}{a_{\rm G}^3}}.
\label{n_G}
\end{eqnarray}

 In Appendix A, the quasi-static tidal torque equations are integrated 
to give $n_{\rm G}(t)$, $n_{\rm bi}(t)$, $\Omega_\ast(t)$, $\Omega_{\rm pp}(t)$ $(=\Omega_{\rm cp}(t))$ as explicit functions of time $t$: 
\begin{eqnarray}
n_{\rm G}(t) &=&
\left[\frac{39}2\frac1{G(M_{\rm pp}+M_{\rm cp})(GM_\ast)^{2/3}} \right. \nonumber \\
 && \hspace{-1.5cm} \times   \left(\frac{k_{\rm 2pp}}{Q_{\rm pp}}R_{\rm pp}^5
\! + \! \frac{k_{\rm 2cp}}{Q_{\rm cp}}R_{\rm cp}^5
\! + \! \frac{M_{\rm pp}^2+M_{\rm cp}^2}{M_\ast^2}\frac{k_{2\ast}}{Q_\ast}R_\ast^5\right)t \nonumber \\
&&  \left. +n_{\rm G,0}^{-13/3}\right]^{-3/13}, \\ 
n_{\rm bi}(t) &=& 
\left[\frac{39}2\frac1{\{G(M_{\rm pp}+M_{\rm cp})\}^{5/3}}\frac1{M_{\rm pp}M_{\rm cp}} \right. \nonumber \\
&& \times \left(\frac{k_{\rm 2pp}}{Q_{\rm pp}}R_{\rm pp}^5M_{\rm cp}^2+\frac{k_{\rm 2cp}}{Q_{\rm cp}}R_{\rm cp}^5M_{\rm pp}^2\right)t \nonumber \\
&& \left. +n_{\rm bi,0}^{-13/3}\right]^{-3/13}, 
\end{eqnarray}
\begin{eqnarray}
\Omega_\ast(t) \!\!\!\! &=&  \!\!\!
-\frac{k_{2\ast}R_\ast^3}{\alpha_\ast Q_\ast}
\frac{M_{\rm pp} \! + \! M_{\rm cp}}{M_\ast } 
\frac{(GM_{\rm pp})^2+(GM_{\rm cp})^2}{(GM_\ast)^{4/3}}  \nonumber \\
& \times & 
 \{n_{\rm G}^{-1/3}(t)-n_{\rm G,0}^{-1/3} \}  \nonumber \\
& \times &
\left[ \frac{k_{\rm 2pp}R_{\rm pp}^5}{Q_{\rm pp}}+\frac{k_{\rm 2cp}R_{\rm cp}^5}{Q_{\rm cp}}  \right. \nonumber \\
&& \left. +\left\{  \left( \frac{M_{\rm pp}}{M_\ast} \right)^2   +  \left( \frac{M_{\rm cp}}{M_\ast} \right)^2  \right\}
\frac{k_{2\ast}R_\ast^5}{Q_\ast} \right]^{-1}  \nonumber \\
& +&  \Omega_{\ast,0}, 
\end{eqnarray}
\begin{eqnarray}
\Omega_{\rm pp}(t) &=&\Omega_{\rm cp}(t) \nonumber \\
= &&  \hspace{-0.8cm}
-\frac{k_{\rm 2pp}R_{\rm pp}^3}{\alpha_{\rm pp} Q_{\rm pp} M_{\rm pp}} 
\left[\frac{(GM_{\rm pp})M_{\rm cp}^3}{\{G(M_{\rm pp}+M_{\rm cp})\}^{1/3}} \right. \nonumber \\
\times && \hspace{-0.8cm}
\left\{ n_{\rm bi}^{-1/3}(t)-n_{\rm bi,0}^{-1/3} \right\} \nonumber \\ 
\times && \hspace{-0.8cm} \left( \frac{M_{\rm cp}^2k_{\rm 2pp}R_{\rm pp}^5}{Q_{\rm pp}}+\frac{M_{\rm pp}^2k_{\rm 2cp}R_{\rm cp}^5}{Q_{\rm cp}} \right)^{-1} \nonumber \\
+&&  \hspace{-0.8cm}
(M_{\rm pp}+M_{\rm cp})(GM_\ast)^{2/3} \{n_{\rm G}^{-1/3}(t)-n_{\rm G,0}^{-1/3} \} \nonumber \\
&&  \hspace{-0.8cm}
\times \left[   \frac{k_{\rm 2pp}R_{\rm pp}^5}{Q_{\rm pp}}+ \frac{k_{\rm 2cp}R_{\rm cp}^5}{Q_{\rm cp}}
\right. \nonumber \\
&&  \hspace{-0.8cm}
+ \left. \left. \left\{ \left( \frac{M_{\rm pp}}{M_\ast} \right)^2 
+ \left( \frac{M_{\rm cp}}{M_\ast} \right )^2 \right\} 
\frac{k_{2\ast}R_\ast^5}{Q_\ast}  \right]^{-1} \right]  \nonumber \\
&&  \hspace{-0.8cm}
+ \Omega_{\rm pp,0}, 
\end{eqnarray}
where the subscripts ",0" represent the values at $t=0$, $k_2$'s are Love numbers,
and $Q$'s are tidal dissipation functions.
We adopt the estimate for the current Solar and Jovian values of $k_2$ and $Q$ as
parameter values of the host star and the planets, which are shown in Table~\ref{alphak}.

\begin{table*}\begin{center}
\begin{tabular}{cccccc}\tableline\tableline
& $\alpha$ & $k_{2}$ & $Q$ & References \\ \tableline
Sun & 0.059 & 0.002 & $10^6$ &  \citet{goldreich,yoder} \\
Jupiter & 0.254 & 0.5 & $10^5$  & \citet{Sasaki} \\ \tableline
\end{tabular}
\caption{
The tidal parameters of the Sun and the Jupiter:
moment of inertia ratios $\alpha$, Love numbers $k_2$,
and tidal dissipation functions $Q$.
\label{alphak}
}\end{center}
\end{table*}

Using these equations, we show in Fig.~\ref{jumyou} the orbital evolutions of binaries at $a_{\rm G}=0.2$ AU (upper panel), 0.3 AU (middle panel), and 0.4 AU (lower panel).
The dash-dotted green and dashed blue lines represent $n_{\rm{bi}}$ and $\Omega_{\rm pp}$.
The vertical red lines represent the lifetime of main sequence phase of solar-type stars ($t=10^{10}$ yr), the upper horizontal black lines are the binary orbital angular velocities for a contact binary
($[G(M_{\rm pp}+M_{\rm cp})/(R_{\rm pp}+R_{\rm cp})^3]^{1/2}$), and the lower one represents $n_{\rm crit}$ that is
determined by the critical semi-major axis of the binary orbit, 
below which the binary separation is so large that the orbit is destabilized by stellar gravitational force
(eq. [\ref{eq:a_crit}]).
When the dash-dotted green lines stay in the region surrounded two horizontal black lines ($n_{\rm crit} < n_{\rm{bi}} < [G(M_{\rm pp}+M_{\rm cp})/(R_{\rm pp}+R_{\rm cp})^3]^{1/2}$ ) the binary planets are stable.
The two planets collide for 
$n_{\rm{bi}} \ge[G(M_{\rm pp}+M_{\rm cp})/(R_{\rm pp}+R_{\rm cp})^3]^{1/2}$ and they escape from each other
for $n_{\rm{bi}} \le n_{\rm crit}$.

The spins of the individual planets are slowed down
by the planet-planet tidal interaction and accordingly
the binary orbital angular momentum is increased.
As a result, both $\Omega_{\rm pp}(t)$ and $n_{\rm{bi}}(t)$ decrease.
Since $\Omega_{\rm pp}(t)$'s deceleration is faster than that of $n_{\rm{bi}}(t)$, $\Omega_{\rm pp}(t)$ catches up with 
$n_{\rm{bi}}(t)$ to establish a synchronous state.
After that, the binary planets keep the synchronized state while $n_{\rm G}(t)$ is decelerated by the tidal torque from the star.
 In this synchronous state with $\Omega_{\rm pp}=\Omega_{\rm cp}=n_{\rm bi}$,
the total angular momentum is given by
\begin{eqnarray}
L \!\! &=& \!\! L_{\rm bi} \!+ \! L_{\rm G} \!+ \! I_\ast\Omega_\ast \! +\! I_{\rm pp}\Omega_{\rm pp}
\! +\! I_{\rm cp}\Omega_{\rm cp} \label{eq:totangular}, \\
& = & \!\! \frac{M_{\rm cp}(GM_{\rm pp})}{\{n_{\rm{bi}}G(M_{\rm pp}+M_{\rm cp})\}^{1/3}}\nonumber\\
&& + \frac{(M_{\rm pp}+M_{\rm cp})(GM_\ast)^{2/3}}{n_{\rm{G}}^{1/3}}
\nonumber \\
&+&
\alpha_\ast R_\ast^2M_\ast\Omega_\ast \nonumber \\
&& +
(\alpha_{\rm pp} R_{\rm pp}^2M_{\rm pp}+
\alpha_{\rm cp} R_{\rm cp}^2M_{\rm cp})n_{\rm bi}, 
\end{eqnarray}
where $\alpha$'s are moment of inertia ratios, the values of which we adopted are shown in Table~\ref{alphak}.
Since $L_{\rm G}$ $(\propto n_{\rm G}(t)^{-1/3})$ increases continuously, 
$\Omega_{\rm pp}(t)$ and $n_{\rm{bi}}(t)$ 
increase to keep the synchronized state 
from the total angular momentum conservation
($L_{\rm bi}$ and $I_\ast\Omega_\ast$ decrease, but $L_{\rm G}$, $I_{\rm pp}\Omega_{\rm pp}$, and $I_{\rm cp}\Omega_{\rm cp}$ increase).
Because $n_{\rm{bi}}(t)$ keeps increasing, $a_{\rm bi}$ keeps decreasing and 
eventually the binary planets collide with each other.

The synchronous state among $\Omega_{\rm pp}$, $\Omega_{\rm cp}$, and $n_{\rm{bi}}$ 
is established in about 0.3 Myr in the cases of $a_{\rm G}=0.3$ AU and 0.4 AU.
However, in the case of $a_{\rm G}=0.2$ AU,
it becomes $n_{\rm bi}<n_{\rm crit}$ and two planets escape from each other 
in about 0.1 Myr before the synchronous state is established.
The lifetime of the binary planets is longer than main-sequence phase of
solar-type stars ($\sim 10$ Gyrs) for $a_{\rm G}=0.4$ AU.
For $a_{\rm G}=0.3$ AU, the lifetime is about 7 Gyrs.
But, since the tidal torque is proportional to fifth-order of the planetary radius (equation [\ref{tauij}]),
the lifetime for $R_{\rm pp},R_{\rm cp}=1R_{\rm J}$ with the same $M_{\rm pp},M_{\rm cp}=1M_{\rm J}$ is lengthened by about 10 times
from that in this plot.
Because gas envelope may fully contract in 0.1 Gyr, $R_{\rm pp}, R_{\rm cp}=1R_{\rm J}$ may be more appropriate than $2R_{\rm J}$ for the estimate of
the binary lifetime.
Thereby, the binary would survive also for $a_{\rm G}=0.3$ AU.

We adopted 10 hours as the initial planetary spin periods, $2\pi /\Omega_{\rm pp,0}$
and $2\pi /\Omega_{\rm cp,0}$.
For smaller $\Omega_{\rm pp,0}$ and $\Omega_{\rm cp,0}$, 
the lifetime of the binary is shorter,
but the lifetime is still $\sim20$ Gyr for $a_{\rm G}=0.4$ AU even if $\Omega_{\rm pp,0}=n_{\rm bi,0}$.
On the other hand,
when the planetary spin period is $\la 4$ hours,
the binary is separated more than $a_{\rm crit}$ before reaching the synchronous state.

 The $Q$ value may include large uncertainty.
However, the binary stability condition for $a_{\rm G}$ comes from
the tidal evolution timescale due to stellar tide compared with 10 Gyrs. 
Since the timescale is proportional to $Q_{\rm pp} a_{\rm G}^{6.5}$ (Eq.~\ref{mitsumori2}), the condition for $a_{\rm G}$ is not severely affected by the uncertainty 
in the $Q$ value.

Note also that we neglect the spins in planet-planet scattering calculations.
The planetary spin axis can be reversed in scattering.
The binary approaches and impacts each other quickly by tidal evolution when binary orbit and spin are retrograde. 
When stellarcentric orbit and binary orbit are retrograde, the 1st, 4th, and 5th
terms in equation (\ref{eq:totangular}) change their signs and the binary is separated away.
That means even if $a_{\rm G}\ga 0.3$ AU, a part of the binary planets cannot survive.

\section{CONCLUSIONS}
\label{kousatsu}

In this paper, we have studied the formation of binary planets 
(a gravitationally bound pair of planets like a planet-satellite system) 
by the capture due to planet-planet {\it dynamical} tide 
during orbital crossing of three giant planets 
and the following long-term evolution due to {\it quasi-static} planet-planet and
planet-star tides.

The scattering of three giant planets usually ends up with ejection, a collision
between planets, or a collision with the central star.
\citet{nagasawa} have found that some fraction of paths to
collisions with the central star can be replaced by formation of a hot jupiter
if planet-star tidal interaction is included.
Here, we have pointed out that some fraction of collisions
between planets are replaced by formation of the binary planets
if planet-planet tidal interaction is incorporated.

Through N-body simulations taking into account 
planet-planet and planet-star tidal interactions (dynamical tide), we have found the followings:
\begin{enumerate}
\item
The binary planets are formed in $\sim10$ \% of
the three planet systems that undergo orbital crossing.
The fraction is independent of stellarcentric orbital radius 
(at least in a range of $0.5 \; \rm{AU}-10 \; \rm{AU}$ that we examined).
Although the formation probability is lower than that found in optimized two planet setting
by \citet{posi}, it is still non-negligible.
Since our initial settings are much more realistic, 
the $10\%$ probability encourages observational detection.

\item
The binary planets tend to be formed in early stage of orbital instability.
In fact, almost all binary planets are formed around their original locations.

\item
Initial captures usually occur at separations of $\sim 1-2$ times of the sum of
planetary radii $R_{\rm tot}=(R_i+R_j)$, resulting in highly eccentric orbits
with a pericenter distance of $\sim (1-2) R_{\rm tot}$ after the trapping.
Through tidal circularization, the pericenter distance expands by a factor of 2
because of the conservation of the binary's angular momentum.
Finally, binary planets with separations of $\sim (2-4)R_{\rm tot}$ are formed.
\end{enumerate}

\vspace{2cm}
 Because dynamical tide diminishes as eccentricity of the binary orbits decreases,
subsequent orbital evolution is dominated by long-term quasi-static tides.
We studied the long-term evolution of the formed binary planets,
taking into account planet-planet and planet star 
quasi-static tidal interactions.
We found that
\begin{enumerate}
\item[4.] If the stellarcentric semimajor axis is larger than 0.3 AU,
the binary is not destroyed during main sequence lifetime of solar type stars
($\sim 10^{10}$ years). 
\end{enumerate}

During the long-term tidal evolution, we have neglected
the effect of a third planet.
It is very rare that the third one enters the Hill sphere of the binary
and the third one hardly affects the binary tidal evolution.
\citet{gong} showed that even if a loosely bounded satellite 
with a separation $\sim 0.1r_{\rm Hill}$ survives strong orbital scattering
by another planet with $\sim20$ \% of probability.

Since we can predict a frequency of binary planets, binary separations and 
a range of stellarcentric semimajor axis where binary planets exist,
we are greatly interested in detectability of extrasolar binary planets.
\citet[][Paper II]{Ochiai} concludes that among various observational methods,
detecting modulations of transit light curves is the most promising.
If radial velocity follow-up can determine the mass of the bodies,
the bulk density derived by assuming a hypothetical single planet
would be $\sqrt2$ times lower than the real bulk density of binary planets.
Thereby, some of objects classified as inflated gas giants or false positives
could be binary planets.  
We will discuss these observation issues in details in Paper II.

%
%
%
%

\acknowledgments

We thank Takahiro Sumi, Karen Lewis, Tristan Guillot and Rosemary Mardelling for discussions on observations of binary planets.
We also thank 
Takayuki Tanigawa and Hidenori Genda for helpful theoretical comments.
This research was supported by a grant for
JSPS (23103005) Grant-in-aid for Scientific Research on Innovative Areas.

\appendix

\section{Long-term evolution due to quasi-static tides}

We calculate the tidal evolution of the star-primary-companion system following \citet{Sasaki}.
The torque exerted on the object $i$ from the object $j$ is given by \citet{SSD} as
\begin{eqnarray}
\tau_{i-j} = -\frac32\frac{Gk_{2i}R_i^5M_j^2}{Q_ia_j^6}{\rm{sgn}}(\Omega_i-n_j),
\label{tauij}
\end{eqnarray}
where $n_j$ is the orbital mean motion of the object $j$ around the object $i$
and $\Omega_i$ is its spin angular velocity.

Here we consider a pair of comparable planets.
But, we call one planet "primary" and the other "companion" for convenience, and
use the subscripts ''$\ast$'' for the central star, ''pp'' for the primary planet, 
and ''cp"'' for the companion planet.
The total angular momentum $L$ is
\begin{equation}
L =
L_{\rm bi}+L_{\rm G}+I_\ast\Omega_\ast+I_{\rm pp}\Omega_{\rm pp}+I_{\rm cp}\Omega_{\rm cp} \label{L}, 
\end{equation}
where $I_i=\alpha_iR_i^2M_i$ is the inertia moment of the object $i$ and
\begin{eqnarray}
\label{L_G}
L_{\rm G} &=& 
(M_{\rm pp}+M_{\rm cp}) \sqrt{G(M_\ast + M_{\rm pp}+M_{\rm cp})a_{\rm G}}
=\frac{(M_{\rm pp}+M_{\rm cp})(GM_\ast)^{2/3}}{n_{\rm G}^{1/3}(t)},\\
L_{\rm bi} &=& 
\frac{M_{\rm pp} M_{\rm cp}}{M_{\rm pp}+M_{\rm cp}}
\sqrt{G(M_{\rm pp}+M_{\rm cp})a_{\rm bi}}
= \frac{M_{\rm cp}(GM_{\rm pp})}{\{G(M_{\rm pp}+M_{\rm cp})n_{\rm bi}(t)\}^{1/3}},
\label{L_bi}
\end{eqnarray}
are the angular momenta of stellarcentric and the binary orbits, respectively 
(where we use equations (\ref{n_bi}) and (\ref{n_G}) and the assumption 5 in \S \ref{chousekishinkanosetsu}).

\citet{Sasaki} neglected the interactions between the central star and the companion planet (in their case, "moon") and their spins.
Since we also consider the binary planets with relatively small
stellarcentric orbital radius where the transit detectability are high,
we include all the tidal interactions and the spin angular momenta in the star-planet-companion system. 
The spin and angular momenta change rates are written as
\begin{eqnarray}
I_\ast\frac{d\Omega_{\ast}}{dt} &=& \tau_{\ast-{\rm pp}}+\tau_{\ast-{\rm cp}}, \\
I_{\rm pp}\frac{d\Omega_{\rm pp}}{dt} &=& \tau_{{\rm pp}-{\rm cp}}+\tau_{{\rm pp}-\ast}, \\
I_{\rm cp}\frac{d\Omega_{\rm cp}}{dt} &=& \tau_{{\rm cp}-{\rm pp}}+\tau_{{\rm cp}-\ast}, \\
\frac{dL_{\rm G}}{dt} &=& -\tau_{\ast-{\rm pp}}-\tau_{{\rm pp}-\ast}-\tau_{\ast-{\rm cp}}-\tau_{{\rm cp}-\ast}, \label{dLGdt} \\
\frac{dL_{\rm bi}}{dt} &=& -\tau_{{\rm pp}-{\rm cp}}-\tau_{{\rm cp}-{\rm pp}},
\end{eqnarray}
where we use the assumption 4 in \S \ref{chousekishinkanosetsu}.

We calculated the tidal evolution using above equations.
From the time derivations of equations (\ref{L_G}) and (\ref{L_bi}), we derived
\begin{eqnarray}
\frac{dn_{\rm G}}{dt} &=&
-\frac3{(M_{\rm pp}+M_{\rm cp})(GM_\ast)^{2/3}}n_{\rm G}^{4/3}(t)\frac{dL_{\rm G}}{dt}, \label{eq:nG}\\
\frac{dn_{\rm bi}}{dt} &=&
-\frac{3\{G(M_{\rm pp}+M_{\rm cp})\}^{1/3}}{M_{\rm cp}(GM_{\rm pp})}n_{\rm bi}^{4/3}(t)\frac{dL_{\rm bi}}{dt},
\end{eqnarray}
and the integrations of these equations gives
\begin{eqnarray}
n_{\rm G}(t) &=&
\left[\frac{39}2\frac1{G(M_{\rm pp}+M_{\rm cp})(GM_\ast)^{2/3}}\left(\frac{k_{\rm 2pp}}{Q_{\rm pp}}R_{\rm pp}^5+\frac{k_{\rm 2cp}}{Q_{\rm cp}}R_{\rm cp}^5+\frac{M_{\rm pp}^2+M_{\rm cp}^2}{M_\ast^2}\frac{k_{2\ast}}{Q_\ast}R_\ast^5\right)t+n_{\rm G,0}^{-13/3}\right]^{-3/13},
\nonumber \\
\\
n_{\rm bi}(t) &=&
\left[\frac{39}2\frac1{\{G(M_{\rm pp}+M_{\rm cp})\}^{5/3}}\frac1{M_{\rm pp}M_{\rm cp}}\left(\frac{k_{\rm 2pp}}{Q_{\rm pp}}R_{\rm pp}^5M_{\rm cp}^2+\frac{k_{\rm 2cp}}{Q_{\rm cp}}R_{\rm cp}^5M_{\rm pp}^2\right)t+n_{\rm bi,0}^{-13/3}\right]^{-3/13},
\nonumber \\
\end{eqnarray}
where the subscripts ",0" represent the values at $t=0$, $k_2$'s are Love numbers,
and $Q$'s are quality $Q$ factors.

The spin angular velocities are derived from the similar methods as
\begin{eqnarray}
\Omega_\ast(t) &=&
-\frac{k_{2\ast}R_\ast^3}{\alpha_\ast Q_\ast}
\frac{M_{\rm pp}+M_{\rm cp}}{M_\ast }
\frac{(GM_{\rm pp})^2+(GM_{\rm cp})^2}{(GM_\ast)^{4/3}} \nonumber \\
& \times &
\frac{ \{n_{\rm G}^{-1/3}(t)-n_{\rm G,0}^{-1/3} \} }{\left[k_{\rm 2pp}R_{\rm pp}^5/Q_{\rm pp}+k_{\rm 2cp}R_{\rm cp}^5/Q_{\rm cp}+\{(M_{\rm pp}/M_\ast)^2+(M_{\rm cp}/M_\ast)^2\}k_{2\ast}R_\ast^5/Q_\ast \right]} \nonumber \\
& + & \Omega_{\ast,0}, \\
\nonumber \\
\Omega_{\rm pp}(t) &=&\Omega_{\rm cp}(t) =
-\frac{k_{\rm 2pp}R_{\rm pp}^3}{\alpha_{\rm pp} Q_{\rm pp} M_{\rm pp}} \nonumber \\
& \times &
\left\{\frac{(GM_{\rm pp})M_{\rm cp}^3}{\{G(M_{\rm pp}+M_{\rm cp})\}^{1/3}}
\frac{ n_{\rm bi}^{-1/3}(t)-n_{\rm bi,0}^{-1/3} }
{\left( M_{\rm cp}^2k_{\rm 2pp}R_{\rm pp}^5/Q_{\rm pp}+M_{\rm pp}^2k_{\rm 2cp}R_{\rm cp}^5/Q_{\rm cp} \right)}
\right. \nonumber \\
&+&
\left.\frac{(M_{\rm pp}+M_{\rm cp})(GM_\ast)^{2/3} \{n_{\rm G}^{-1/3}(t)-n_{\rm G,0}^{-1/3} \} }{\left[ k_{\rm 2pp}R_{\rm pp}^5/Q_{\rm pp}+k_{\rm 2cp}R_{\rm cp}^5/Q_{\rm cp}+\left\{ (M_{\rm pp}/M_\ast)^2+(M_{\rm cp}/M_\ast)^2 \right\} k_{2\ast}R_\ast^5/Q_\ast \right] }\right\} \nonumber \\
 & +& \Omega_{\rm pp,0}.
\end{eqnarray}
When the spins of the binary planets become synchronous with the binary orbital rotation, we can write $\Omega_{\rm pp}=\Omega_{\rm cp}=n_{\rm bi}$.
From the total angular momentum conservation,
\begin{eqnarray}
L(t \ge \tau_1) &=&
\frac{M_{\rm cp}(GM_{\rm pp})}{\{n_{\rm{bi}}(t)G(M_{\rm pp}+M_{\rm cp})\}^{1/3}}+
\frac{(M_{\rm pp}+M_{\rm cp})(GM_\ast)^{2/3}}{n_{\rm{G}}^{1/3}(t)}
\nonumber \\
&+&
\alpha_\ast R_\ast^2M_\ast\Omega_\ast(t)+
(\alpha_{\rm pp} R_{\rm pp}^2M_{\rm pp}+
\alpha_{\rm cp} R_{\rm cp}^2M_{\rm cp})n_{\rm bi}(t),
\label{LgeT1}
\end{eqnarray}
where $t=\tau_1$ is the time when the synchronous state begins.

However, the binary planets become unstable if the binary separation becomes 
large enough before they are tidally rocked.
The critical semi-major axis of the binary orbit ($a_{\rm crit}$),
beyond which the binary orbit is destabilized by stellar gravitational force is
\begin{eqnarray}
a_{\rm crit} = f r_{\rm Hill} = f \left(\frac{M_{\rm pp}+M_{\rm cp}}{3M_\ast}\right)^{1/3}a_{\rm G}.
\label{eq:a_crit}
\end{eqnarray}
We take $f=0.36$ following \citet{Sasaki}.
The binary orbit becomes unstable, when
\begin{eqnarray}
n_{\rm bi}(t)<n_{\rm crit} \equiv \sqrt{\frac{G(M_{\rm pp}+M_{\rm cp})}{a_{\rm crit}^3}}.
\end{eqnarray}

The timescale of the binary's tidal evolution is
\begin{eqnarray}
\tau_1 &\sim& \frac{n_{\rm bi}}{dn_{\rm bi}/dt}
=-\frac13 \frac{L_{\rm bi}}{dL_{\rm bi}/dt}
= \frac19 \frac1{\sqrt{2GM_{\rm pp}}}\frac{Q_{\rm pp}}{k_{\rm 2pp}R_{\rm pp}^5}a_{\rm bi}^{6.5} \nonumber \\
&\sim& 10^5\bun{a_{\rm bi}}{5R_{\rm pp}}^{6.5} \;\; \rm{yr},
\label{mitsumori}
\end{eqnarray}
where the physical parameters of the planets are the same.
Taking the major term of equation (\ref{dLGdt}),
the timescale of tidal evolution of binary's barycenter $\tau_2$ is derived by the similar method as
\begin{eqnarray}
\tau_2 &\sim& \frac{n_{\rm G}}{dn_{\rm G}/dt}
=-\frac13 \frac{L_{\rm G}}{dL_{\rm G}/dt}
\sim \frac29 \frac1{\sqrt{GM_\ast}}\frac{M_{\rm pp}}{M_\ast}\frac{Q_{\rm pp}}{k_{\rm 2pp}R_{\rm pp}^5}a_{\rm G}^{6.5} \nonumber \\
&\sim& 10^{16}\bun{M_\ast}{M_\odot}^{-1.5}\bun{M_{\rm pp}}{M_{\rm J}}^{-2/3}\bun{a_{\rm G}}{1\rm{AU}}^{6.5} \;\; \rm{yr},
\label{mitsumori2}
\end{eqnarray}
which is much longer than the estimation of equation (\ref{mitsumori}).


\begin{thebibliography}{}
\bibitem[Beaug\'{e} \& Nesvorn\'{y}(2012)]{BN12}Beaug\'{e}, C. \& Nesvorn\'{y}, D. 2012, ApJ, 751, 119
\bibitem[Chambers et al.(1996)]{Chambers}Chambers, J. E., Wetherill, G. W., \& Boss, A. P. 1996, Icarus, 119, 261
\bibitem[Chatterjee et al.(2008)]{c08}Chatterjee, S., Ford, E. B., Matsumura, S., \& Rasio, F. A. 2008, ApJ, 686, 580
\bibitem[Ford \& Rasio(2008)]{FR08}Ford, E. B. \& Rasio, F. A. 2008, ApJ, 686, 621
\bibitem[Gladman(1993)]{Gladman}Gladman, B. 1993, Icarus, 106, 247
\bibitem[Goldreich \& Soter(1966)]{goldreich}Goldreich, P. \& Soter, S. 1966, Icarus, 5, 375
\bibitem[Gong et al.(2013)]{gong}Gong, Y.-X. Zhou, J.-L. Xie, J.-W., \& Wu, X. M. 2013, ApJL, 769, L14
\bibitem[Ikoma et al.(2006)]{Iko06} Ikoma, T., Guillot, T., Genda, H., Tanigawa, T., \& Ida, S. 2006, ApJ, 650, 1150
\bibitem[Juri\'c \& Tremaine(2008)]{Juric08}Juri\'c, M. \& Tremaine, S. 2008, \apj, 686, 603
\bibitem[Lega et al.(2013)]{Lega13}Lega, E., Morbidelli, A., \& Nesvorn\'y D. 2013, MNRAS, 431, 3494
\bibitem[Lin \& Ida(1997)]{LI97}Lin, D. N. C \& Ida, S. 1997, ApJ, 477, 781
\bibitem[Mardling(1995)]{Mardling}Mardling, R. A. 1995, ApJ, 450, 732
\bibitem[Marzari \& Weidenschiling(2002)]{MW02}Marzari, F. \& Weidenschiling, S. J. 2002, Icarus, 156, 570
\bibitem[Murray \& Dermott(1999)]{SSD}Murray, C. D. \& Dermott, S. F. 1999, Solar System Dynamics (Cambridge: Cambridge Univ. Press)
\bibitem[Nagasawa et al.(2008)]{nagasawa}Nagasawa, M., Ida, S., \& Bessho, T. 2008, ApJ, 678, 498
\bibitem[Nagasawa \& Ida(2011)]{NI11}Nagasawa, M. \& Ida, S. 2011, ApJ, 742, 72
\bibitem[Ochiai et al.(2014)]{Ochiai}Ochiai, H., Lewis, K. M., Nagasawa, M., \& Ida, S. 2014, submitted (Paper II)
\bibitem[Podsiadlowski et al.(2010)]{posi}Podsiadlowski, P., Rappaport, S., Fregeau J. M., \& Mardling, R. A. 2010, arXiv:1007.1418
\bibitem[Portegies Zwart \& Meinen(1993)]{zm}Portegies Zwart, S. F. \& Meinen, A. T. 1993, A\&A, 280, 174
\bibitem[Press \& Teukolsky(1977)]{PT77}Press, W. H. \& Teukolsky, S. A. 1977, ApJ, 213, 183
\bibitem[Rasio \& Ford(1996)]{RF96}Rasio, F. A. \& Ford, E. B. 1996, Science, 274, 954
\bibitem[Sasaki et al.(2012)]{Sasaki}Sasaki, T., Barnes, J. W., \& O'Brien, D. P. 2012, ApJ, 754, 51
\bibitem[Weidenschilling \& Marzari(1996)]{Weiden96}
Weidenschilling, S. J. \& Marzari, F., 1996, Nature, 384, 619
\bibitem[Yoder(1995)]{yoder}Yoder, C. F. 1995, Astrometric and Geodetic Properties of Earth and the Solar System, (Washington, D.C.: American Geophysical Union)
\end{thebibliography}
\end{document}